\documentclass{llncs}
\usepackage{llncsdoc}

\usepackage{rotating} % for sidewaystable

\usepackage{psfrag}
\newcommand{\argmax}{\mathop{\mathrm{argmax}}}

\usepackage[cmex10]{amsmath}
\usepackage{bbm}
\usepackage[caption=false,font=footnotesize]{subfig}

\usepackage{graphicx}
\usepackage{amssymb}

\usepackage{wrapfig}
\usepackage{sidecap}

\usepackage{flushend}

\begin{document}
\markboth{\LaTeXe{} Class for Lecture Notes in Computer
Science}{\LaTeXe{} Class for Lecture Notes in Computer Science}

\title{\vspace*{-1.3cm} Contextual Multi-armed Bandits for the Prevention of Spam in VoIP Networks}

\author{ Technical Report \\ \medskip Tobias Jung\inst{1} \and Sylvain Martin\inst{1} \and Damien Ernst\inst{1} \and Guy Leduc\inst{1}\\
{\tt \{tjung,sylvain.martin,dernst,guy.leduc\}@ulg.ac.be}}

\institute{Montefiore Institute, University of Li\`ege, Belgium}

\maketitle

\begin{abstract}
In this paper we argue that contextual multi-armed bandit algorithms
could open avenues for designing self-learning security modules for 
computer networks and related tasks. The paper has two contributions:
a conceptual one and an algorithmical one. The conceptual contribution
is to formulate -- as an example -- the real-world problem of preventing SPIT (Spam in 
VoIP networks), which is currently not satisfyingly addressed by standard
techniques, as a sequential learning problem, namely as a contextual 
multi-armed bandit. Our second contribution is to present CMABFAS, 
a new algorithm for general contextual multi-armed bandit learning that
specifically targets domains with finite actions. We illustrate how CMABFAS
could be used to design a fully self-learning SPIT filter that does not
rely on feedback from the end-user (i.e., does not require labeled data)
and report first simulation results.
\end{abstract}

%==================================================================
\section{Introduction}
\label{seq:introduction}
%==================================================================
SPIT is an acronym for spam in internet telephony and refers to unsolicited calls 
that, when answered by a human, would deliver a pre-recorded message (e.g., advertisement or phishing attempts). 
Similar to spam in emails, SPIT exploits the openness of the existing infrastructure (e.g., no 
strongly authenticated identities) together with the fact that VoIP calls can be easily generated 
automatically and at zero (or very low) costs. Unlike with spam in emails however, where the content consists
of text and can be analyzed before it is delivered, the content of a phone call (a voice stream) is only 
available when the call is answered. Thus many of the defensive measures that are effective against email spam 
do not directly translate to SPIT mitigation.
Previously, some first ideas have already been suggested to address this problem. 
They range from reputation-based  \cite{Kolan07,fde} and call-frequency based \cite{Shin06} dynamic black-listing, 
fingerprinting \cite{C06}, to challenging suspicious calls by captchas \cite{schlegel06,dtmf-captcha-for-SPIT,quitt:icc}, or 
the use of standard machine learning such as anomaly detection \cite{nassar-SVM,nassar10}, clustering \cite{wu09},
or decision trees \cite{sin11}. 
We believe that with respect to SPIT prevention these earlier solutions suffer from one or both of the following
two shortcomings: (1) they are built on weak ``features'' (i.e., information from the protocol header SIP which in
essence are text strings produced by the VoIP client) which are fairly easy to manipulate for a sophisticated
hacker; (2) they are built as a {\em static} defense from labeled training data (e.g., signatures of known attacks),
require constantly manual adjustments from a human domain expert and are thus vulnerable to novel
attacks.

In this paper we explore a novel paradigm in machine learning, namely reinforcement learning, to attack 
this problem from a new angle. Specifically, we use {\em contextual multi-armed bandits} to design
a {\em self-learning} SPIT filter which dynamically selects from among several security 
policies (i.e., voice captchas and computational puzzles) the most appropriate one and which
does not rely on explicit feedback from the end-user (i.e., labeled training data)---instead, the SPIT filter
monitors its own performance and generates internal rewards from subsequent traffic.
 
The paper is structured as follows. We begin in Section~2 with describing the philosophy behind the design
of our SPIT filter and motivate the use of contextual bandits to implement it in the real world. The
following Section~3 introduces our algorithm CMABFAS, a variant of a contextual MAB for learning in finite
action spaces over generalized metric spaces, which will have exactly the properties we need for our 
SPIT filter. Note that our description of CMABFAS in this section will be kept general (such that it can 
be applied to other problems as well); in Section~4 we then describe in detail how we can map the 
SPIT prevention problem from Section~2 to CMABFAS. Section~5 then presents some simulation results and 
compares CMABFAS with a more na\"{\i}ve baseline implementation of MABs.

%==================================================================
\section{Background and related work}
%==================================================================
{\em Before we can start, we feel it necessary to give a fair warning to the reader.} At the present time, 
VoIP telephony has not (yet) replaced traditional telephony and the problem of SPIT is largely a
hypothetical one. In particular, there is no publicly available dataset\footnote{The earlier work 
described in \cite{bab} set out to precisely change that. In it the authors describe a methodology for creating SPIT traffic 
and also provide a common data set for the use in benchmark comparisons. However, the data set they provide is generated 
from ``emulated users based on a social model''; in essence, the authors use common tools to generate the SPIT traffic,
where the relevant features, such as call duration, inter-arrival time, behavior upon receiving a call, etc. are all modeled by 
sampling from distributions. For example, the call duration was generated from an exponential distribution the parameter of which
was specified by hand (which amounts to the same as what we do here). 
%In other words: the dataset does not contain any ``real'' spit if by ``real'' we mean 
%either ``received by a real human'' or ``sent by an unknown third party with the intent of making 
%the callee buy something''.
} and little experience of what SPIT will look like in the real world. To the best of our knowledge, 
the ``empirical evaluation'' of all earlier research on SPIT prevention is therefore based on guesswork and simulation. 
In this paper, we will face the same situation; however, our method also has to {\em interact}
with the calling party -- which is even harder to simulate realistically and cannot be done from a static dataset. 
Our method will therefore also be evaluated in only a simplified testbed where the behavior of SPIT bots is ``emulated'' 
by distributions the parameters of which are synthetically chosen by hand.

%==================================================================
\subsection{Related work: Designing a SPIT filter}
%==================================================================
We consider the {\em inbound} scenario, where the SPIT filter is located close to the recipient of the call 
(e.g., at the VoIP proxy) and decisions must be made on a per-call basis without having explicit history information 
on a per-source basis (thus ruling out reputation-based methods). For every incoming
call the system has to decide whether it is a regular call or a SPIT call using only features directly extracted 
from the request, i.e., text strings extracted from the fields of the Session Initiation Protocol (SIP) 
header in an INVITE message.
Current hardware phones are identifiable through the specific SIP header they produce,
and it is conceivable that SPIT bots could be equally identifiable through the specific SIP header they produce.
Nevertheless we believe that, from this information alone, traditional {\em static} techniques are not sufficient to build a 
strong filter that detects and blocks SPIT with high accuracy. This has two reasons: (1) any such signature-based
defense will require a human expert to manually identify and add signatures of SPIT bots to a list of known attacks
which is a costly procedure and leaves the system vulnerable to novel attacks; and (2) the information in the 
protocol header is weak in the sense that it can be easily manipulated by a sophisticated hacker to make a 
SPIT call appear as if originating from a regular device.

An interesting idea suggested in \cite{schlegel06,gritzalis2008136} and on which we are going to build is for the defensive system to collect additional
information that would be a lot harder for SPIT to manipulate; so-called Turing tests or 
voice CAPTCHAS that would {\em actively interact} with the calling party. For example, before 
forwarding a call, an automated mechanism could prompt a suspicious caller to dial a short sequence of
randomly generated digits. Both the reaction of the caller to the test (a bot is not likely to obey
telephone etiquette and would immediately start to play back its pre-recorded message), as well
as the result of the test itself (only a very sophisticated bot will be able to break a voice CAPTCHA)
will reveal additional information about whether or not the caller is a human or a SPIT bot.
A large number of these security challenges, most of which
are parameterized to generate an infinite variety, already exists today; 
however, deciding which of these security challenges to best apply given 
the features of a call is currently done by a human expert (e.g., see NEC's SEAL \cite{schlegel06}).  
Deciding for which call what security challenge to best apply is, however, not trivial. On the one side, 
applying a challenge will reveal additional information about the call being SPIT or not. 
On the other side, applying it will also carry certain costs, namely: (1) annoying the calling party; (2)
additional computational resources; and (3) obfuscation, meaning that we would prefer to
avoid exposing all capabilities of our defense system such that attackers can not start to learn
from them. The essential point here is that, while it would result in the least number of mistakes, 
we cannot afford to apply our strongest but likely most ``costly'' security challenge to every single call. 

Based on this design for a SPIT filter which can choose from many possible security challenges or {\em actions} 
(where we include ``apply no security challenge'' as just another action), our goal is to create
a {\em self-learning} SPIT filter which does not rely on hand-coded rules but automatically determines from
past experience what the ``best'' security challenge should be for a given call.  This self-learning does 
not rely on external feedback; instead the system monitors its own performance and generates internal 
rewards. Moreover, this self-learning also ensures that the system will adapt to new variants of SPIT as part of its 
normal operation.

\begin{wrapfigure}{R}{0.32\textwidth}
\vspace{-20pt}
\begin{center}
\includegraphics[width=0.32\textwidth]{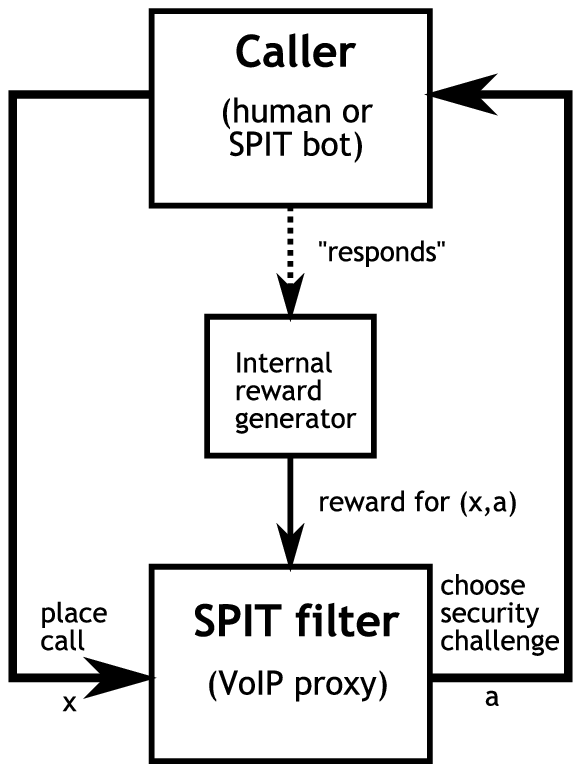}
\end{center}
\vspace{-20pt}
\end{wrapfigure}
A sketch of the basic interaction loop between caller and SPIT filter is shown to the right. In this figure,
calls are processed sequentially (individually one by one). Every time a call arrives at the SPIT filter, 
the SPIT filter selects one action and applies it to the call. Each action forces a response from the caller
(e.g., passing/failing a security challenge or, if the call is made, features from the call such as call
duration, amount of double talk, etc.). This response is analyzed by the internal reward generator by first
inferring whether or not the caller is a SPIT bot. Then, depending on the outcome of this inference stage
(which may be a probability for the call being SPIT), the nature of the action chosen (if it is likely SPIT,
did we chose an action that tried to prevent it), and the cost of the action, a scalar reward is returned to
the SPIT filter. From this reward the SPIT filter updates its internal (call,action)-scores and proceeds
with the next call in the queue. 

In summary, to implement this design for a SPIT filter, we have to address two issues: (1) how to implement
the reward generator; (2) how to implement the action selection and learning part. Note that in this paper
we focus on the latter.

%==================================================================
\subsection{Background: Multi-armed bandits}
\label{seq:MAB}
%==================================================================
To implement learning, we formalize our SPIT filter as a 
multi-armed bandit problem (MAB) with context. {\em Standard MABs} are well-studied
models for sequential decision-making when the outcome is stochastic and its distribution
a priori unknown. In a standard MAB we assume we get to play the following ``game'' over
multiple rounds: suppose we are given $n$ different choices or actions and each action
is associated with a stochastic reward function (that stays the same for all rounds we play
but may be different for each action). In every round of the game we have to choose one
of the $n$ possible actions, and in doing so we obtain a random reward sampled from 
the underlying distribution. Our goal is to choose actions such that the 
{\em sum of rewards} we obtain is maximized.

Naturally, the best action would be to always choose the action yielding the highest 
expected reward. However, the reward distributions are not revealed to the player and
thus it is (initially) unknown which action will produce the highest reward. To solve this
problem, we have to form an estimate for each action about what reward we might get, based 
on what results we have obtained in earlier rounds of the game. Of course, the more often
we have tried a particular action in the past, the more certain and reliable this estimate
will be. The fundamental dilemma is now how to balance exploitation (choosing what 
currently appears to be the best action) and exploration (choosing a non-greedy
action to improve our estimate and potentially obtain higher rewards in the future).
The standard MAB problem with finite (and small) number of actions is largely
considered to be a solved problem and provable optimal strategies exist in the
literature \cite{AA,AB}.

A {\em contextual MAB} is in principle like a standard MAB with the major difference that it
is defined over some large set of elements or a continuous space. Now each element
of the set corresponds to a separate standard MAB (in some formulations also 
the action space becomes large or continuous). Learning the reward distributions in 
contextual MABs is more challenging than it is for standard MAB since it is no longer 
possible to sample the same action multiple times. Instead, we have to 
impose ``smoothness'' as additional structural assumption; i.e., we assume that 
elements of the context space that are ``similar'' (with respect to some similarity measure) 
will also behave similarly. Using
generalization we can then try to predict the outcome for new cases based on previously
observed outcomes for similar cases. Contextual MAB are nowadays an active research topic
with many relevant real-world applications, e.g., placement of web advertisements. See 
\cite{b1,b2,b3,DBLP:journals/corr/abs-0907-3986,Kleinberg_STOC08,b4} for some examples.

We believe that contextual MABs (but not standard MABs) are a good description of what the 
SPIT filter motivated in the previous section is trying to achieve: the contexts correspond 
to calls (represented by SIP headers), the actions correspond to security challenges 
the filter can choose from, and the rewards correspond to how the calling party reacts.

%==================================================================
\section{Description of CMABFAS}
%==================================================================
This section describes
our algorithm CMABFAS for contextual MAB in finite action spaces.
Note that we keep the presentation general, the actual application 
to the SPIT prevention scenario will be described in the following
Section~\ref{sec:section5}. Our work is largely based on the contextual 
zooming algorithm described in \cite{DBLP:journals/corr/abs-0907-3986},
and inspired by the X-armed bandit learning algorithm described in \cite{Xarmed_bubeck}. 
Differences between CMABFAS and \cite{DBLP:journals/corr/abs-0907-3986} are: a specialization to 
the finite action case and a modified scheme to estimate the expected rewards which works for
more general metric spaces (we do not need the triangle inequality). 

%==================================================================
\subsection{Notation} 
%================================================================== 
We begin by introducing some notation. Let $\mathcal X$ denote the context space 
with elements $x \in \mathcal X$ and let $a \in \{1,\ldots,k\}$ denote the possible
actions that can be chosen for each $x$ (we assume that we have the same choice of 
actions available for each $x$). Each context $x$ can be seen as an index to a conventional $k$-armed bandit: 
for each $x$ we have a distribution of rewards $\mathcal R^a(x)$ under action
$a$ which models the stochastic response from the environment when performing
action $a$ in context $x$. Let $r^a(x)$ denote the random values drawn from the
corresponding reward distribution, i.e., $r^a(x)\sim \mathcal R^a(x)$. We assume that
$\mathcal R^a(x)$ has bounded support which, for notational convenience, is 
supposed to lie in the unit interval; thus we assume 
$\mathrm{supp} \, \mathcal R^a(x) \subseteq [0,1]$. In consequence, we also have
$r^a(x) \in [0,1]$. 

Let $\mu^a(x)$ denote the mean of the reward distribution $\mathcal R^a(x)$; i.e.,
\begin{equation}
\mu^a(x):= E_{r^a(x)\sim \mathcal R^a(x)} \bigl[r^a(x)\bigr]. 
\end{equation}

The essence of the problem we study is that $\mathcal R^a(x)$, and hence
$\mu^a(x)$, is not known when making decisions; instead it is treated as a
black-box from which only samples can be drawn. Our overall goal is, when 
presented with any context $x$, to be able to choose the action with the 
highest mean: $\argmax_a \, \mu^a(x)$. The algorithm we propose is 
based on taking and averaging samples in a smart way such that observations
we made at one location $x$ are reused to make an estimate about the
reward distribution at another location $x'$.

Let context space $\mathcal X$ be equipped with a distance metric 
$d(x,x')$ which measures the distance between any two elements 
$x,x'\in \mathcal X$. We assume that $d$ is a pseudo-metric and fulfills the
conditions of (1) non-negativity: $d(x,x')\ge 0$; (2) symmetry: 
$d(x,x')=d(x',x)$; and (3) is equal to zero if and only if the arguments
are identical: $d(x,x')=0 \Leftrightarrow x=x'$, $\forall x,x'\in\mathcal X$.  
Furthermore we assume that, for notational convenience again, $d$ is scaled
such that the diameter of $\mathcal X$ with respect to $d$ is equal to 
one; that is, $\sup_{x,x'\in \mathcal X} \, d(x,x')=1$. 

Finally, to make learning and generalization over the context space at all 
possible, we need to impose smoothness and restrict the variation of the
mean functions $\mu^a$. Specifically, we assume that each $\mu^a$ is 
Lipschitz with a modulus of variation $\lambda>0$:
\begin{equation}
|\mu^a(x)-\mu^a(x')| < \lambda d(x,x'), \quad \forall x,x'\in \mathcal X, \forall a.
\label{eq:lipschitz}
\end{equation}

%==================================================================
\subsection{Objective} 
%==================================================================
The goal of contextual MAB is to find a strategy for the following ``game''. 
The game proceeds in rounds $t=1,2,\ldots$. At each round $t$
we observe $x_t \in \mathcal X$; we suppose that we have no control over how
$x_t$ is generated from the set $\mathcal X$ and furthermore that the 
mechanics leading to the selection of an $x_t$ are independent from 
whatever happened in all previous rounds of the game. Given $x_t$, we have
to choose an action $a_t \in \{1,\ldots,k\}$. Executing action $a_t$
lets us then observe the reward $r^{a_t}(x_t)$ which is a random sample 
from $\mathcal R^{a_t}(x_t)$. Our goal is to use the results of the $t-1$
previous rounds of the game, i.e., the history 
\[(x_1,a_1,r^{a_1}(x_1),\ldots,x_{t-1},a_{t-1},r^{a_{t-1}}(x_{t-1})),\]
to determine an action $a_t$ such that the regret -- a measure of performance --
is minimized.   
In the bandit literature two types of  regret are considered:
here we take the cumulative regret which assumes that we have to play the game 
for a fixed number $T$ of rounds
and that we want to minimize over all $T$ rounds the difference between
the expected reward of the best possible action minus
the expected reward of the action chosen at that round:
\begin{equation}
\textrm{regret}=\sum_{t=1}^T |\mu^*(x_t)-\mu^{a_t}(x_t)|, 
\label{eq:regret}
\end{equation} 
where $\mu^*(x_t)=\max_{a \in \{1,\ldots,k\}} \, \mu^a(x_t)$.

%==================================================================
\subsection{Illustration} 
%==================================================================
\psfrag{R1x}{\tiny$\mathcal R^1(x'')$}
\psfrag{R2x}{\tiny$\mathcal R^2(x')$}
\psfrag{mu1x}{\scriptsize$\mu^1(x)$}
\psfrag{mu2x}{\scriptsize$\mu^2(x)$}
\psfrag{m2}{\tiny$\mu^1(x'')$}
\psfrag{m1}{\tiny$\mu^2(x')$}
\psfrag{x}{\tiny$x'$}
\psfrag{x'}{\tiny$x''$}
\psfrag{X}{\scriptsize$\mathcal X$}
\psfrag{reward}{\scriptsize reward}
\psfrag{samples1}{\tiny$r^2(x')$}
\psfrag{samples2}{\tiny$r^1(x'')$}
\psfrag{action 1}{\tiny action 1}
\psfrag{action 2}{\tiny action 2}
\psfrag{optimal}{\tiny optimal}

\psfrag{samples11}{\tiny =samples $r^1(\cdot)\sim \mathcal R^1(\cdot)$}
\psfrag{samples22}{\tiny =samples $r^2(\cdot)\sim \mathcal R^2(\cdot)$}
\psfrag{xt?}{\scriptsize$x_t=?$}

\begin{figure}[!t]
\begin{center}
		\includegraphics[width=3.6in]{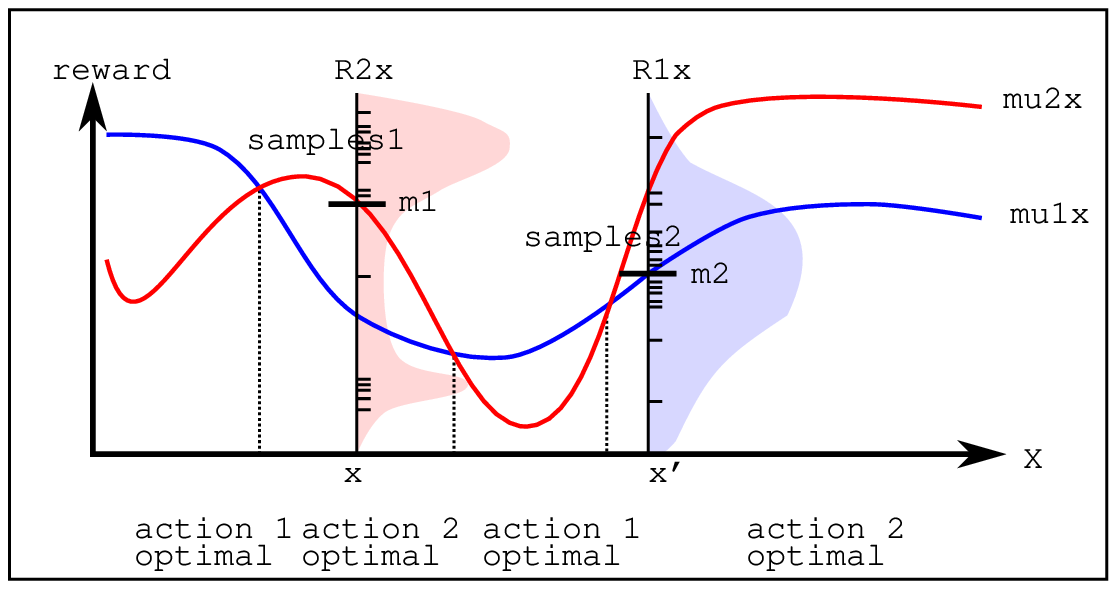}	\\ \medskip
			
		\includegraphics[width=3.6in]{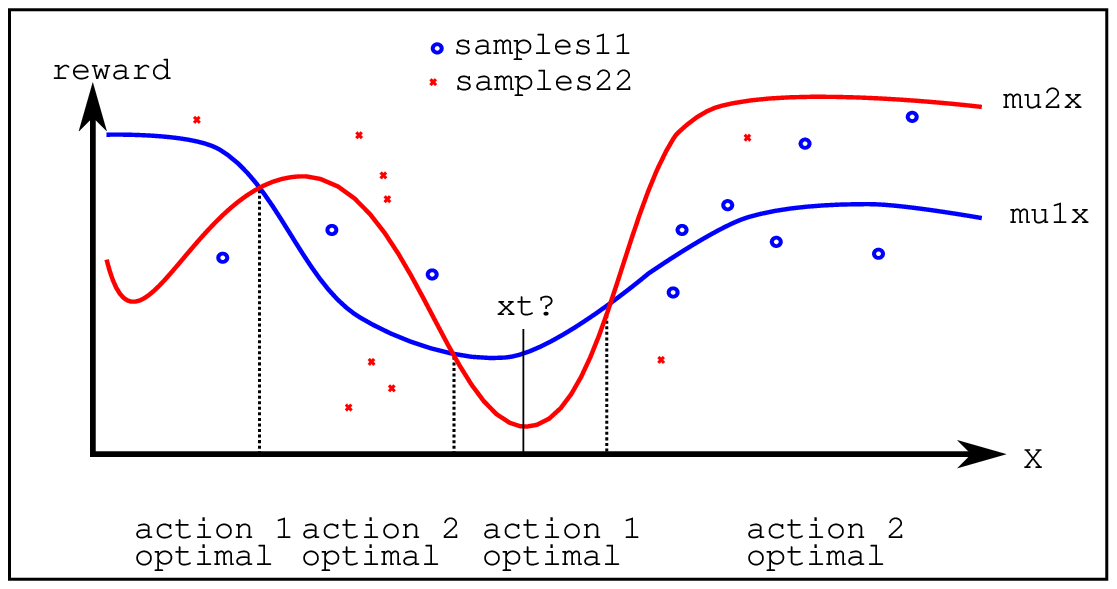}
\end{center}		
\caption{Contextual multi-armed bandit with two actions (blue) and (red) over context 
space $\mathcal X$ (in our case $\mathcal X$ will be the space of SIP headers). 
Each point/location $x \in \mathcal X$ is associated with an action-dependent reward 
distribution the mean of which is denoted by the blue and red curve. Top: sample rewards are
observed at location $x'$ and $x''$ with the shaded area denoting the underlying distribution.
Bottom: a more realistic illustration; in practice, multiple reward samples are rarely obtained 
at the same location. Instead they are spread out and need to be aggregated in a judicious way
to estimate the expected reward at a new query location $x_t$.}
\label{fig:drawing}
\end{figure}
Figure~\ref{fig:drawing} tries to depict the whole situation graphically. In it, we chose to
draw $\mathcal X$ as a straight line, suggesting that it is a continuous and
compact space (however in theory it can equally well be a discrete or finite
space). Mean reward $\mu^a(x)$ is drawn as a continuous function over $\mathcal X$;
the figure shows two mean functions $\mu^1(x)$ and $\mu^2(x)$ corresponding
to two possible actions $a \in \{1,2\}$. For each location $x$ and action $a$
we have a separate reward distribution $\mathcal R^a(x)$ from which 
random samples $r^a(x)$ are gathered.

The goal is to find for each $x$ the action with the highest expected reward, 
which in our illustration is the curve which is on top of the other curve. As
the figure shows, in general we will not have the situation that one and the
same action is optimal everywhere. Instead, because of the smoothness\footnote{Note 
that in our mathematical formulation of the problem smoothness is only imposed
over the means of the distribution. The actual form of the distribution (such as
being concentrated around the mean, being multi-modal, etc.) could vary
from location to location and thus also impact the practical performance.} 
assumptions we made for $\mu^a$, there will be ``regions'' where one action 
is optimal, and regions where another one is optimal.

Note that the figure is somewhat misleading in the way the random samples are 
shown; from Figure~\ref{fig:drawing}(top) it appears as if multiple samples from the same distribution
(i.e., same location and same action) can be gathered. This however is exactly
the situation we do not have (it would correspond to the traditional 
multi-armed bandit scenario). A more realistic representation of the situation we
face is thus Figure~\ref{fig:drawing}(bottom); it shows how the samples are spread out over 
different locations and actions and motivates why at all it becomes necessary to 
average and generalize over the context space $\mathcal X$.

%==================================================================
\subsection{CMABFAS -- High-level overview} 
\begin{figure}[!t]
\begin{center}
\includegraphics[width=0.95\textwidth]{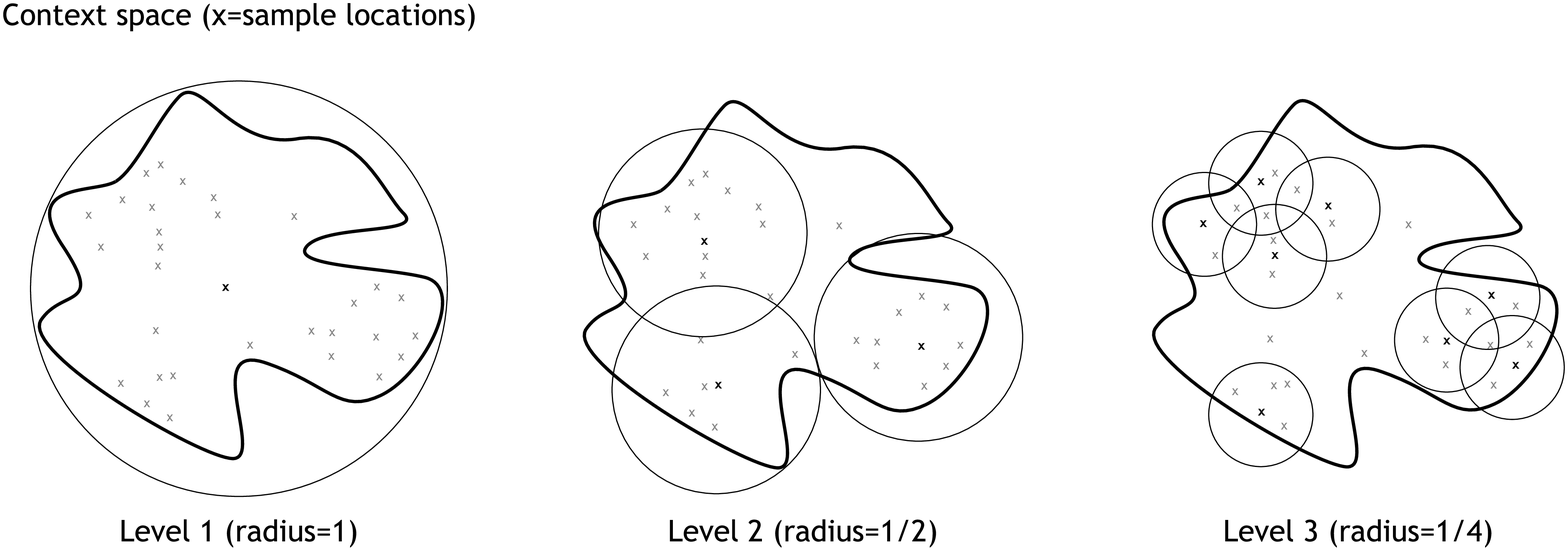}
\end{center}
\caption{Adaptively covering the context space with ball-shaped regions (see text)}
\label{fig:12}
\end{figure}

%==================================================================
Our algorithm CMABFAS works as follows. For each action $a$ separately, we incrementally
construct over time $t=1,2,\ldots$ a cover of the context space $\mathcal X$. The cover 
consists of ball-shaped regions where individual balls are centered on a certain subset
of the contexts $\{x_1,\ldots,x_{t-1}\}$ seen so far. The cover is hierarchical with the radius
of the balls exponentially fast decreasing with the level of hierarchy; e.g., $\mathcal X$ is covered 
at level $1$ by a single ball of radius $1$, at level $2$ by balls of radius $1/2$, at 
level $3$ by balls of radius $1/4$, and so on (see Figure~\ref{fig:12}). Each ball aggregates the reward samples
lying within: we only store their number and their sum. Each ball covering $x_t$ can thus
be used to estimate the expected reward $\mu^a(x_t)$ at query point $x_t$. Since we have to 
balance exploration and exploitation, we will augment this estimate by a UCB-like term (i.e., 
an upper confidence bound). Each ball
covering $x_t$ thus gives rise to a score which is composed of two parts: (1) the sample 
average within the ball, and (2) an uncertainty term which depends on the number
of samples (the fewer samples we have in a ball, the less certain we can be
about the correctness of their average) and the volume of the ball (the larger 
the region the ball covers, the more variation of $\mu^a$ is possible and thus the 
more samples of a different base quantity are lumped together). The ``best ball'' 
for each action is the one with the lowest score (the tightest upper bound), and 
among them the highest score indicates the best action.  
The cover is adaptively refined by adding new balls according to the following rules: 
(1) a new ball can only be created centered at $(x_t,a_t)$; (2) a new ball can 
only be created if the number of samples in the parent ball exceeds a certain
threshold (which grows inverse quadratically in the radius of the parent ball);
(3) a new ball is only created if it will not overlap with already existing balls
at the same level in the hierarchy. Informally speaking, CMABFAS works by ensuring
that high resolutions are only attained in regions of $\mathcal X$ where the 
corresponding action is optimal and then exploiting that balls with increasingly smaller
radius provide increasingly more accurate bounds for the expected reward.

From a practical point of view our algorithm possesses two properties which make it
ideally suited for operation under heavy-load real-world conditions: (1) it is an
{\em online} algorithm whose computational complexity and storage requirements are
very low (this is so because we do not have to store and operate on an always growing
number of individual data points, but only have to store and operate on balls of a fixed radius, 
which is a much smaller
number\footnote{The number of balls of radius $r$ is trivially upper bound by the $r$-packing number of
$\mathcal X$. A more tighter bound can be achieved by the near-optimality dimension or related concepts 
(e.g., see \cite{Xarmed_bubeck,DBLP:journals/corr/abs-0907-3986}) which also take into account 
the specifics of the actual problem, i.e., $\mu^a$.} and where search operations can be efficiently implemented by appropriate
space partitioning methods such as cover trees \cite{covertree}; (2) it is an {\em anytime}
algorithm which aims at producing the best possible solution in each step of its 
operation and does not need any kind of prior learning phase with data or tuning
to start producing meaningful results.

%==================================================================
\subsection{\bf CMABFAS -- Notation} 
\label{sec:cmab_notation}
%==================================================================
Before we come to the details of the algorithm, we need to introduce some more
notation. 
Let $B_i^a \in \{1,\ldots,n_B^a\}$ be the index of the $i$-th ball and 
$n_B^a$ the total number of balls in the cover for action $a$. Let 
$x(B_i^a)\in \mathcal X$ denote the location of ball $B_i^a$, that is,
the location of its center in $\mathcal X$, and let $r(B_i^a)\in [0,1]$
denote its radius. We say that an element $x \in \mathcal X$ lies in ball
$B_i^a$, written as $x \in B_i^a$, if $d(x,x(B_i^a))\le r(B_i^a)$. 
Overloading the notation, we can identify with $B^i_a$ also the region 
$B_i^a=\{x\in \mathcal X \,| \, d(x,x(B^a_i)) \le r(B_i^a)\}\subset \mathcal X$.
Let $n_t(B_i^a)$ denote the number of all the samples gathered up to time $t$
lying in $B_i^a$, and let $\varrho_t(B_i^a)$ denote their corresponding
sum of rewards. Let 
\begin{align*}
\text{avg}_t (B_i^a) & :=  \varrho_t(B_i^a)/ n_t(B_i^a) \\
\text{conf}_t (B_i^a) & := c \cdot \sqrt{ \log T /n_t(B_i^a)} \\
\text{size} (B_i^a) & :=  2\lambda r(B_i^a)
\end{align*} 
where $c$ is a constant (which in practice will become a tunable parameter of the algorithm).
We say that ball $B_i^a$ is {\em full} (able to spawn a child), whenever 
$\text{conf}_t (B_i^a)<r(B_i^a)$.

%==================================================================
\subsection{CMABFAS -- Algorithm} 
\label{sec:algo_implementation}
%==================================================================
{\bf \noindent Initialization:} At time $t=0$ we initialize the individual cover for each action with a single ball: 
we create a ball centered on an arbitrary element $x \in \mathcal X$ and set its radius to $1$
(such that it covers the whole space):
\begin{align*}
\forall a=1,\ldots,k: & \ \ \text{create } B_1^a \text{ with} \\
x(B_1^a) & := \text{any element of } \mathcal X \\
r(B_1^a) & :=1, \ n_0(B_1^a):=0, \ \varrho_0(B_1^a):=0. 
\end{align*}

{\bf \noindent Every step:} Now suppose that at time $t$ $x_t$ arrives. For each action $a$ separately, we first
compute the indices of those balls that contain $x_t$, which we will call {\em active
balls} for $x_t$: $A^a(x_t):=\{B_i^a| x_t \in B_i^a \}$. From the set of active balls we then 
compute the set of {\em relevant balls} which consists of all balls $B_i^a \in A^a(x_t)$ which
are either not full or allow the creation of a child (with radius $\frac{1}{2}r(B_i^a)$) centered on $x_t$
such that it does not overlap (distance at least $\frac{1}{2}r(B_i^a)$) with an already existing 
ball at this level of the hierarchy: $R^a(x_t):=\{B_i^a\in A^a(x_t) \, | \,\text{conf}_t(B_i^a)>r(B_i^a) 
\vee \nexists B_j^a \in A^a(x_t):r(B_j^a)=\frac{1}{2}r(B_i^a)\}$.
For each ball in $R^a(x_t)$ we then compute its current score, which is a 
high-probability upper bound\footnote{
The argument 
goes as follows. Let $x_t$ be the current context and take any active ball $B_i^a\in A^a(x_t)$.
Let $S_n$ be the set of indices of previous samples lying in the ball and $|S_n|=n$ be their
number. Applying Azuma-Hoeffding for martingales with bounded increments together with the 
union bound, one can show that
\[
P\left(\left|\frac{1}{n} \sum_{s\in S_n} [r^a(x_s)-\mu^a(x_s)]\right| < c\cdot \sqrt{\frac{\log T}{n}}\right) \ \ge \ 1-T^{-2}.
\]  
Using the Lipschitz assumption \eqref{eq:lipschitz} together with the fact that both $x_t$ and
$x_s$, $\forall s$, lie in the same ball $B_i^a$ with radius $r(B_i^a)$ we then have that
\[
|\mu^a(x_s)-\mu^a(x_t)|\le \lambda\cdot d(x_s,x_t) \le \lambda 2r(B_i^a)  \qquad \forall s
\]
and thus $\mu^a(x_s)\le\mu^a(x_t)+\lambda 2r(B_i^a)$. Substituting $\mu^a(x_s)$ accordingly gives
us a lower bound for the left side inside $P(\cdot)$, and, noting that 
$\frac{1}{n} \sum_{s\in S_n} r^a(x_s)=\frac{\varrho_t(B_i^a}{n_t(B_i^a)}$, we obtain
as claimed
\[
P\left(\left|\frac{\varrho_t(B_i^a)}{n_t(B_i^a)}-\mu^a(x_t)\right| < c \cdot \sqrt{\frac{\log T}{n_t(B_i^a)}} + 2\lambda r(B_i^a) \right) \ \ge \ 1-T^{-2}.
\]}  
for the error we make when we use the current in-ball sample average as a proxy for the true
but unknown expected reward $\mu^a(x_t)$. We take the minimum over all the upper bounds as 
score $u(x_t,a)$ for the action in question (i.e., the tightest upper bound):
\begin{equation}
u(x_t,a):= \min_{B_i^a \in R^a(x_t)} \ \bigl[\text{avg}_t(B_i^a)+
\text{conf}_t(B_i^a)+\text{size}_t(B_i^a) \bigr].
\label{eq:ucb}
\end{equation}
Having such a $u$-score for each possible action $a$, we then choose the action which
achieves the highest $u$-score:
\begin{equation}
a^*_t:=\argmax_{a=1\ldots k} \, u(x_t,a).
\label{eq:bestaction} 
\end{equation}
The system executes action $a^*_t$ and observes the stochastic outcome $r^{a^*_t}(x_t)$
which is drawn from the unknown distribution $\mathcal R^{a^*_t}(x_t)$. We then use this
new observation to update all the balls that were active for the action chosen:
\begin{align*}
\forall B_i^a\in A^{a^*_t}(x_t): \ \ \varrho_{t+1}(B_i^a) & = \varrho_t(B_i^a)+r^{a^*_t}(x_t)\\
 n_{t+1}(B_i^a) & = n_t(B_i^a)+1 \\
 \text{all remaining balls:} \ \ \varrho_{t+1}(B_i^a) & = \varrho_t(B_i^a) \\
 n_{t+1}(B_i^a) & = n_t(B_i^a).
\end{align*}

{\bf \noindent Adaptive refinement:} Having determined $a^*_t$ and before updating, we test if 
the ball $B^*$ achieving the minimum in \eqref{eq:ucb} for action $a^*_t$ is full
(and thus allowed to spawn a child). If it is full, we add a new ball $B^*_{\text{child}}$
with center $x_t$ and radius $\frac{1}{2}r(B^*)$ and add its index to the list of active balls.

%==================================================================
\section{CMABFAS for SPIT}
\label{sec:section5}
%==================================================================
This section describes how we can map our original problem, detecting and preventing
SPIT calls (as described in Section~2), to the general self-learning 
decision-making framework CMABFAS described in the previous section. 
This is done as follows:

%==================================================================
\subsection{Defining the context space}
%==================================================================
The context space $\mathcal X$ is chosen to be the space of all possible VoIP calls,
which we represent by the information contained in the SIP header. 
Specifically, we extract the fields source IP addresses, contact information for caller, 
callees and optional \emph{vias}, plus fields that have a phone-specific value such as 
user-agent string, preferred codec and source port number. The SIP addresses of both 
parties are further split into user and host names. The result is combined to form a 
vector of 16 text strings. 
For example, one such call $x\in \mathcal X$ is of the form
\begin{align*} 
x = &[\text{\small``208.51.215.203'',``193.22.119.20'',``5838565'',}\\
    &\text{\small``208.51.215.203'',``193.22.119.20'',``5838565'',}\\
    & \text{\small``208.51.215.203'',	``87008888'',	``208.51.215.203'',}\\
    & \text{\small``87008888'',``Cisco-SIPGateway/IOS-12.x'',	``208.51.215.203'',}\\
    & \text{\small``CiscoSystemsSIP-GW-UserAgent'', ``	208.51.215.203'',}\\
    & \text{\small``18660	G723/8000''}]
\end{align*}
To measure distances in $\mathcal X$, we define a metric over SIP headers in 
the following way:
\[
d(x,x')=\begin{cases}0 &, \text{if } \text{count}(x,x')=16 \\
2^{-\text{count}(x,x')} &, \text{otherwise} \end{cases}. 
\]
where $\text{count}(x,x')$ computes the Hamming distance and
returns the number of string attributes that are identical in $x$ and $x'$ 
({\em count} performs a string comparison for each attribute individually). 
As an example, under this metric $d$ two calls 
have distance $d(x,x')=0$ if each attribute in $x$ is identical to its counterpart in $x'$,
distance $d(x,x')=\frac{1}{2}$ if only 1 attribute in $x$ and $x'$ agrees, and distance
$d(x,x')=1$ if no attribute in $x$ and $x'$ agrees, that is, $x$ and $x'$ are 
completely different. Note that with this definition of $d$ the normalization
requirement $\mathrm{diam}(\mathcal X)=1$ is fulfilled.

%==================================================================
\subsection{Defining the actions}
%==================================================================
The action the system has to decide about consists of choosing which particular
security test out of many possible ones to apply to a given call.  
We assume that both human and automated bot will either pass or fail to pass a security test {\em with a certain probability}. 
In general there will be different types of security
tests with each of them being of a certain difficulty and thus inducing different 
probabilities for success and costs. In our experiments we simplify this setting and define
initially two abstract security tests which we call Type-1 and Type-2. For each security test
and call $x \in \mathcal X$ we assign synthetic success probabilities which we design in such
a way that different kinds of bots exist having each different capabilities to bypass
a particular security test (as will be explained in more detail below).
Overall, the system has the following three actions at its disposal:
\begin{itemize}
\item {\bf A1:} Apply no security test and directly pass the call on to the recipient

\item {\bf A2:} Apply security test Type-1. If the caller is able to pass the test
successfully, forward the call to the recipient. If the caller is not able to pass 
the test successfully in one attempt, flag the call as SPIT.

\item{\bf A3:} Apply security test Type-2 and proceed as in A2.
\end{itemize}

%==================================================================
\subsection{Defining the rewards}
%==================================================================
\begin{figure*}[!t]
\begin{center}
\begin{minipage}{0.3\textwidth}
\begin{center}
{\small
\begin{tabular}{|l||c|c|c|}
\hline
& \bf A1 & \bf A2 & \bf A3 \\
\hline
\hline
normal   & 1.0 & 0.9 & 0.8 \\
honeypot & 1.0 & 0.5 & 0.3 \\
voipbot  & 1.0 & 0.3 & 0.5 \\
warvox   & 1.0 & 0.1 & 0.3 \\
spitter  & 1.0 & 0.3 & 0.3 \\
\hline
\end{tabular}
}
\smallskip

\centerline{\scriptsize (a) Chosen success probs}
\end{center}
\end{minipage}
\begin{minipage}{0.3\textwidth}
\begin{center}
{\small
\begin{tabular}{|l||c|c|c|}
\hline
& \bf A1 & \bf A2 & \bf A3 \\
\hline
\hline
normal   & \bf 120 & 28     & 36 \\
honeypot & 30      & 15     & \bf 49 \\
voipbot  & 30      & \bf 49 & 15 \\
warvox   & 30      & \bf 83 & 49 \\
spitter  & 30      & 49     & \bf 83 \\
\hline
\end{tabular}
}
\smallskip

\centerline{\scriptsize (b) Resulting expected reward}
\end{center}
\end{minipage}
\begin{minipage}{0.3\textwidth}
\begin{center}
{\small
\begin{tabular}{|l||c|c|c|}
\hline
& \bf A1 & \bf A2 & \bf A3 \\
\hline
\hline
normal   & \bf 0 & 92     & 84 \\
honeypot & 19    & 34     & \bf 0 \\
voipbot  & 19    & \bf 0  & 34 \\
warvox   & 53    & \bf 0  & 34 \\
spitter  & 53    & 34     & \bf 0 \\
\hline
\end{tabular}
}
\smallskip

\centerline{\scriptsize (c) Resulting regret}
\end{center}
\end{minipage}

\caption{Reward specification (see text). The optimal action is shown in bold face.}
\label{fig:reward_specification}
\end{center}
\end{figure*}
Defining the rewards is the one rather difficult modeling choice we face. The reward
is a single scalar quantity that must capture the performance of the SPIT filter. It
has to be defined in such a way that by choosing actions which optimize it (which is what
CMABFAS does), the SPIT filter does what we, as its designer, want it to do.
In our case here the reward has to account for two things: (1) did we make the right decision
in letting through a call or rejecting it; and (2) how ``expensive'' were the security
tests necessary to arrive at this decision. While we imagine that the latter can be designed
by a human expert without too much trouble, the first item poses a serious conceptual
challenge: whether or not a call $x$ is SPIT is beyond the system to detect
on its own and cannot be established during runtime. Instead it would require human
feedback much the same as an email spam filter requires labeled data or humans moving 
suspicious email to a dedicated spam folder. However, we would rather like our SPIT filter to be
able to detect {\em by itself} if a call is SPIT (and thus generate the internal reward 
appropriately) without relying on humans pushing a red button every time SPIT gets through.

Motivated by earlier work \cite{SPRIT-SPIT}, we believe that one property\footnote{Of course, other features
would also be possible, e.g., amount of double-talk, time-to-speech etc. (see \cite{SPRIT-SPIT}),
and it is an open question of how to use these features to design rewards properly. Our modeling here should merely be 
seen as a first concept of proof.} of calls which can be used
in this regard is {\em call duration}. The basic idea is that,
on the average, SPIT calls will tend to be of shorter duration than NON-SPIT calls. 
%This is so because humans are likely to hang up on a caller as soon as they realize it is an
% unwanted advertisement. Call duration is a quantity that is automatically recorded 
%by the service provider (e.g., for billing purposes) and thus does not require additional 
%machinery to be installed.  Also, unlike other statistics, raw call duration is harder 
%to manipulate by SPIT bots since it involves human interaction and cannot be artificially
%inflated. 
Based on a large data set of collected real-world call durations \cite{eagle09realitydata},
we will model call duration by an exponential distribution with mean
30 seconds for SPIT calls and mean 120 seconds for NON-SPIT calls. 
If, however, a call is flagged as SPIT, we do not observe its call duration
since the call is not physically answered. in this case we assign it a fixed 
reward of +100. The rationale for assigning +100 is that, on the
average (i) SPIT will fail the security test and NON-SPIT will pass it and (ii) the reward
of +100 is larger than the expected call duration for SPIT and smaller than the expected call
duration for NON-SPIT, thus making one of the actions A2 or A3 optimal for SPIT and action 
A1 always optimal for NON-SPIT.
Finally, whenever we choose action A2 or A3 we always incur, regardless of the outcome, a fixed
cost which was set to -100. In summary, the reward is generated according to the following
rule:
\begin{itemize}
 \footnotesize
 \item Generating the reward for applying action $A1$ to call $x$:
     \begin{itemize}
        \item if $x \in $ SPIT, reward $r^{A1}(x)$ is sampled from $Exp(30)$
        \item if $x \in $ NON-SPIT, reward $r^{A1}(x)$ is sampled from $Exp(120)$ 
     \end{itemize}
 \item Generating the reward for applying action $A2/A3$ to call $x$:
     \begin{itemize}
        \item if $x$ passes the security test (the probability of which depends on $x$ and $A2/A3$)
        \begin{itemize}
            \item if $x \in $ SPIT, reward $r^{A2/A3}(x)$ is sampled from $Exp(30)$ minus cost\_of\_A2/A3
            \item if $x \in $ NON-SPIT, reward $r^{A2/A3}(x)$ is sampled from $Exp(120)$ minus cost\_of\_A2/A3
        \end{itemize}
        \item else if $x$ fails the security test
        \begin{itemize}
            \item reward $r^{A2/A3}(x):=100$ minus cost\_of\_A2/A3 
        \end{itemize}
    \end{itemize} 
\end{itemize}
To model if a call $x$ is able to pass a security test, we generate a single Bernoulli trial whose mean
depends on $x$ and A2/A3 (see Figure~\ref{fig:reward_specification}a).

%==================================================================
\subsection{Setting up the experiment}
%================================================================== 
Finally, we have to discuss how $x$ is related to being SPIT or NON-SPIT. 
Our experiments are based on a dataset which is a capture of 5609 calls from 
a real network operator under non-disclosure agreement. Neither SPIT nor 
other undesired activity was reported during this period of time. Additionally, we 
generated 2827 calls using available security testing tools in a test-bed environment 
and recorded the corresponding SIP messages. Internally, $x$ thus belongs to one of 5 
classes: {\em normal}, {\em warvox} ({\tt http://warvox.org}), {\em spitter} ({\tt http://hackingvoip.com/\\sec\_tools.html}), 
{\em voipbot} ({\tt http://voipbot.gforge.inria.fr}), or {\em honeypot} \\({\tt http://artemisa.sourceforge.net)}. 
The first class is NON-SPIT, the last corresponds to unsolicited scanning activity, and 
the other 3 remaining classes are all different kinds of SPIT with a different signature. 
In our experiment, we assume that each of these classes has different capabilities of passing a given 
security test, making it necessary to combat each class with a different optimal action. These different
capabilities are implemented by assigning (by hand) different success probabilites to each class for 
each action. The reward distributions that results from our choices are summarized in Figure~\ref{fig:reward_specification}. 
Note again that all these detailed mechanics are not known by the CMABFAS SPIT filter; the only thing the filter sees are 
the rewards sampled from the rule given above.

%==================================================================
\section{Simulation Results}
\label{sec:results}
%==================================================================

%==================================================================
\paragraph{Setup}
%==================================================================
To populate the context space, we first generate a corpus of 8436 SIP headers as described
in the previous section (5609 of type normal, 870 of type spitter, 6 of type honeypot, 80 of type 
warvox, 1861 of type voipbot). Every time we simulate a new call, we draw a random header uniformly 
from this corpus and present it to the CMABFAS learner. 
We consider three scenarios of increasing difficulty: one where the SPIT filter has to choose among
three different actions A1,A2,A3, one where it has to choose among 10 different actions A1,$\ldots$,A10,
and one where it has to choose among 50 different actions A1,$\ldots$,A50. The three action scenario
was described in the previous section, the other scenarios are obtained by just adding more choices of
security tests to the disposal of the SPIT filter and setting success probabilities accordingly. Note
that increasing the number of choices makes the task of finding the best choice more difficult (in
that it requires more exploration).
The stochastic rewards are scaled to lie in 
$[0,1]$ and are generated as shown in Figure~\ref{fig:reward_specification} (actions A4$\ldots$A50 are populated similarly). 
The Lipschitz constant $\lambda$ from Eq.~\eqref{eq:lipschitz} is set to $1$. We perform a total of 10 independent runs and 
average the results; each single run consists of sequentially processing 10,000,000 independent calls.

%==================================================================
\paragraph{Baseline}
%==================================================================
To properly evaluate the performance our algorithm CMABFAS, we define a na\"{\i}ve baseline method
which works by incrementally (but non-adaptively) clustering the input space $\mathcal X$
and assigning a standard MAB with UCB$_1$(1.2) rule \cite{UCB1.2} to each cluster. Specifically,
it works like this: let $x_t$ be the current call. Find the nearest cluster according to the Hamming distance.
If the distance to the nearest cluster is greater than some parameter {\tt max\_radius} and the 
number of current clusters is below some other parameter {\tt max\_clusters}, we add a new
cluster and initialize its counter to zero. Otherwise we assign $x_t$ to the nearest cluster 
with index $i^*$ and choose action $a^*$ such that
\[
a^*=\argmax_a \ \frac{\varrho_t(i^*,a)}{n_t(i^*,a)} + \sqrt{\frac{1.2\log(n_t(i^*,a))}{n_t(i^*)}},
\]
where $\varrho_t(i^*,a)$ is the sum of rewards for action $a$, $n_t(i^*,a)$ the number of samples
for action $a$, and $n_t(i^*)$ the total number of samples within cluster $i^*$. Choosing $a^*$, we 
observe, as before, a reward $r^{a^*}(x_t)$ after which we increment $\varrho_t(i^*,a^*)$, $n_t(i^*,a^*)$,
and $n_t(i^*)$ accordingly. The hyperparameters of the algorithm, {\tt max\_radius} and {\tt max\_clusters},
we chosen by a coarse grid search: best performance was achieved for {\tt max\_radius}=6 and 
{\tt max\_clusters}=500 (our results also include some other combinations.)

\begin{figure}[!t]
\centerline{
\includegraphics[width=2.3in]{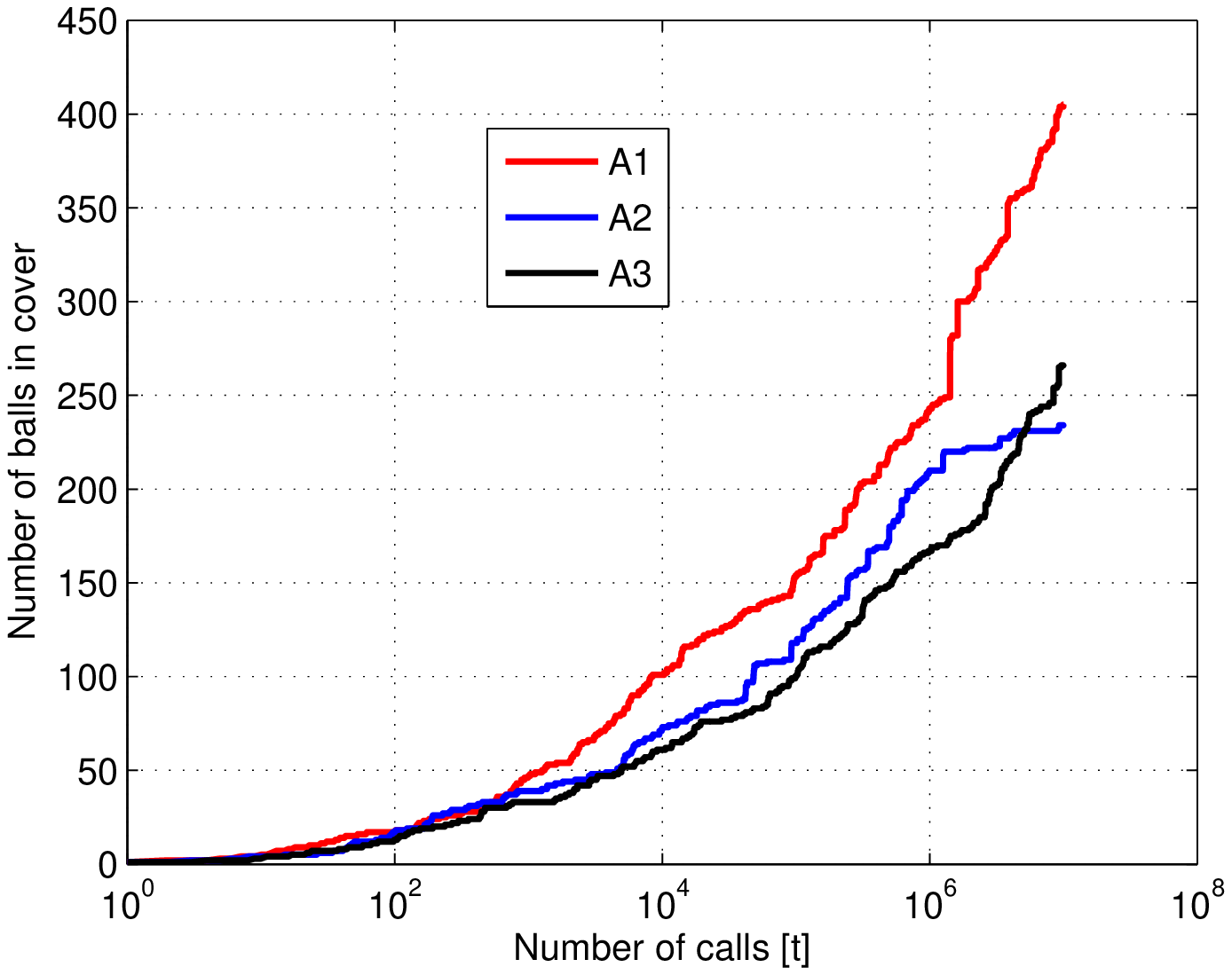}
\includegraphics[width=2.3in]{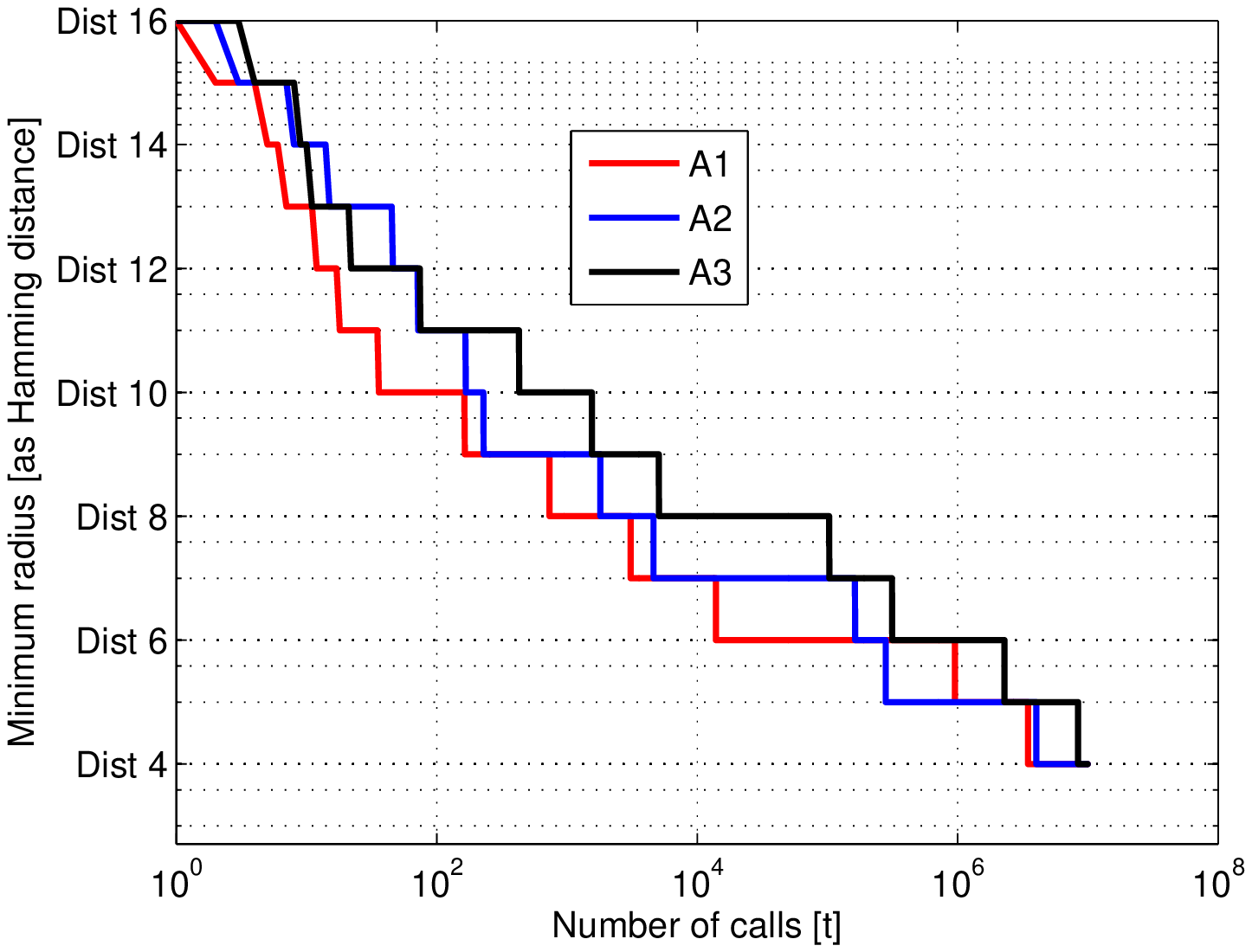}
}
\caption{Left: number of balls CMABFAS creates over time (3 actions). 
Right: How the minimum radius changes over time (3 actions) }
\label{fig:nballs}
\end{figure}

\begin{figure}[!t]
\begin{center}
\begin{minipage}{0.32\textwidth}
\begin{center}
		\includegraphics[width=4.4cm]{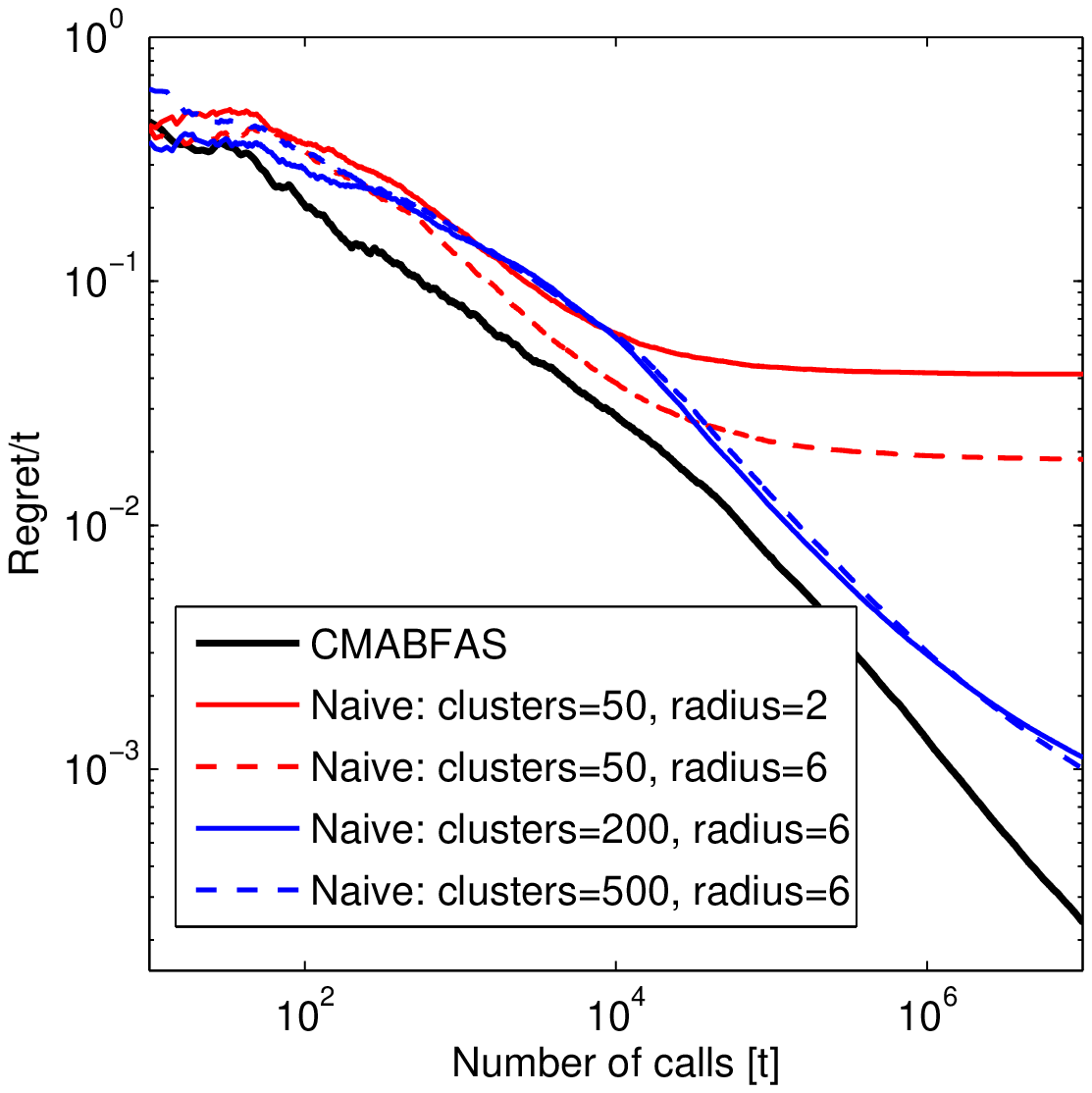}\\
	
	{\small 3 available actions}
	\end{center}
\end{minipage}
\begin{minipage}{0.32\textwidth}
\begin{center}
		\includegraphics[width=4.4cm]{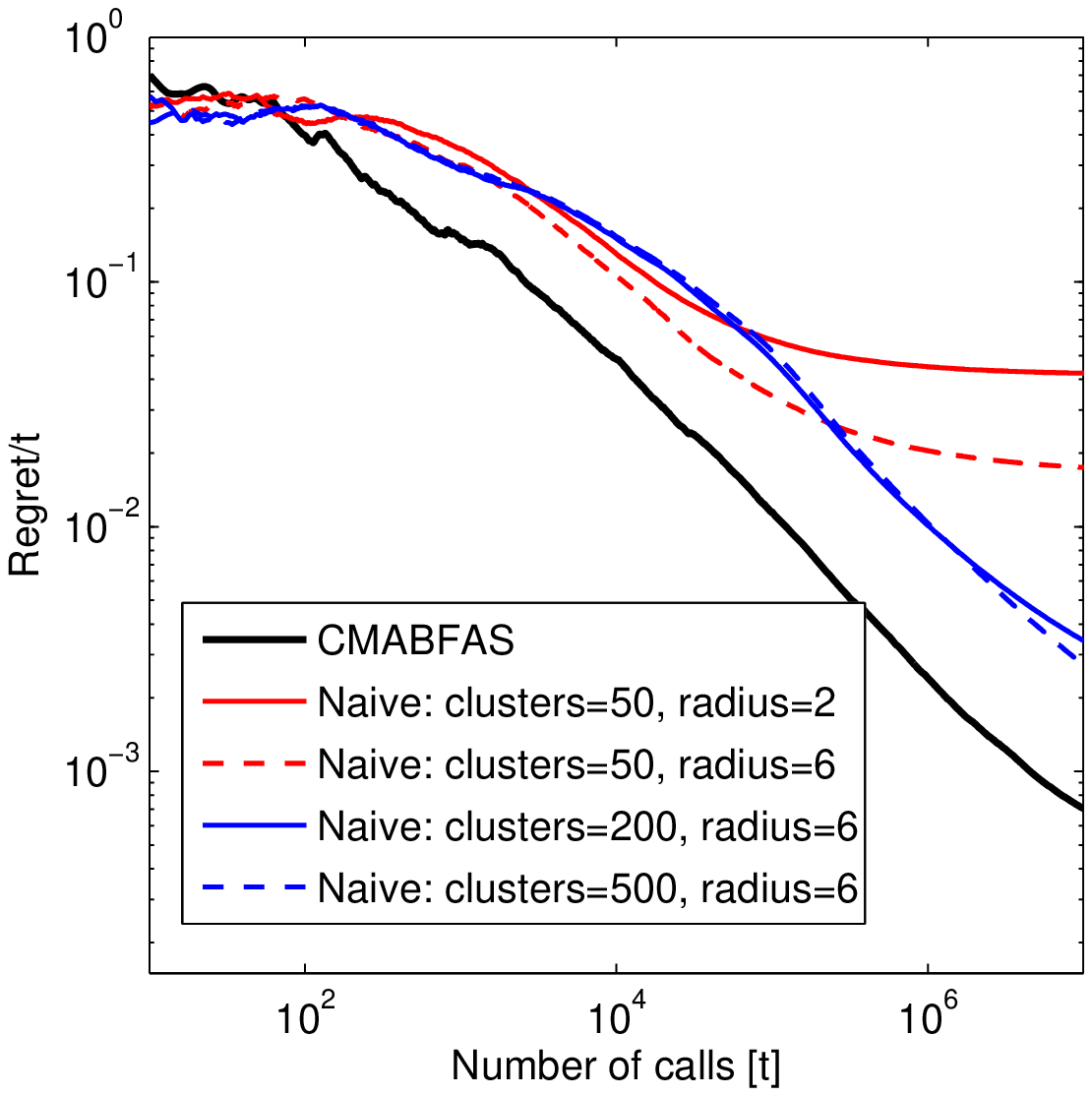}\\
		
		{\small 10 available actions}
\end{center}
\end{minipage}
\begin{minipage}{0.32\textwidth}
\begin{center}
		\includegraphics[width=4.4cm]{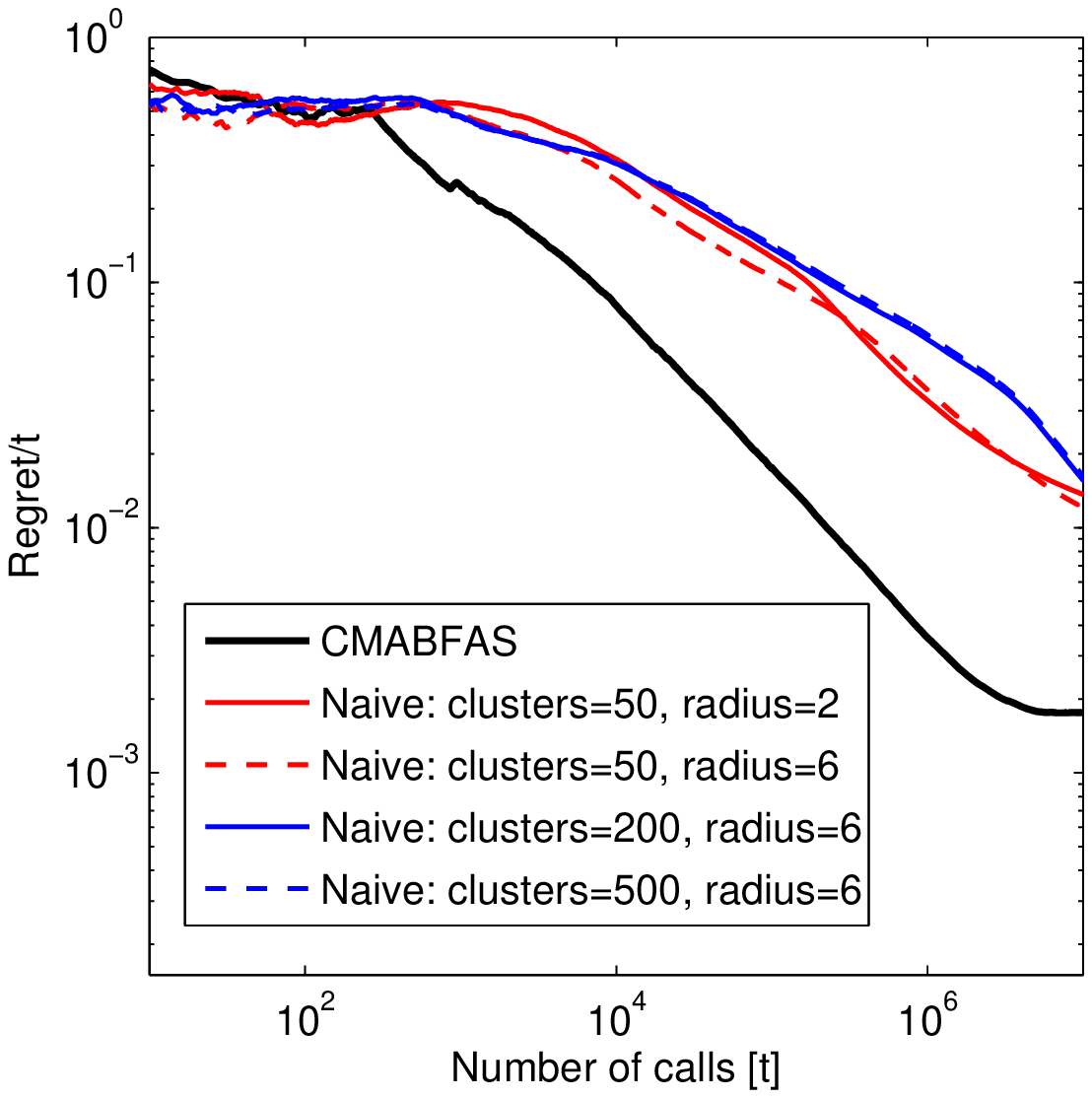}\\
		
		{\small 50 available actions}
\end{center}		
\end{minipage}

\end{center}
\caption{
 Comparing CMABFAS with na\"{\i}ve clustering and standard UCB$_1$ MAB. Lower numbers indicate better performance.}
\label{fig:results}
\end{figure}

%==================================================================
\paragraph{Results}
%==================================================================
The resulting performance of both CMABFAS and the baseline is shown in Figure~\ref{fig:results} in terms 
of the cumulative regret, while Table~\ref{tab:results} shows the results numerically in greater depth. 
Figure~\ref{fig:nballs} illustrates the partitioning behavior of CMABFAS over time. In summary, the results
show that CMABFAS is about an order of magnitude better than the best parameter setting of the baseline. 
The curves reflect the kind of learning behavior that we would have expected and which is typical for MAB
algorithms of this kind: initially, the reward distributions are unknown and thus the algorithm has to ``explore'' and try out the various 
actions over different regions in the context space. Over time, and this happens very rapidly with CMABFAS, 
the granularity of the ball-cover of the context space is refined in areas of high data-density 
and the estimates for the mean reward become more accurate, this in turn makes the algorithm become more
confident about its decisions and explore less. The performance CMABFAS reaches at the end is nearly 
optimal: both the regret and the number of mistakes approach zero (averaged over all calls, the algorithm 
makes the correct decision with $> 99.9$\%). We also note that CMABFAS appears to scale well when the 
number of available actions is increased (see the 10 action and 50 action results).
Finally, recall that CMABFAS is an anytime algorithm. If we 
would continue to run it and process additional calls, the error rate would further decrease until 
(asymptotically) no more mistakes are made.

\begin{sidewaystable}
\begin{center}
\begin{tabular}{|l|rrr|rrr|rrr|}
\hline
& \multicolumn{3}{|c|}{$t=$10,000 calls} &\multicolumn{3}{|c|}{$t=$100,000 calls} &\multicolumn{3}{|c|}{$t=$10,000,000 calls}\\
                                     & regret/t & nmistakes1 & nmistakes2 & regret/t & nmistakes1 & nmistakes2 & regret/t & nmistakes1 & nmistakes2 \\
\hline
Actions 3                            &               &           &           &               &            &            &               &            &           \\
\hspace*{0.5cm}CMABFAS                & {\bf 0.02811} & {\bf 753} & {\bf 374} & {\bf 0.00737} & {\bf 2195} & {\bf 1030} & {\bf 0.00023} & {\bf 6983} & {\bf 1378}\\
\hspace*{0.5cm}Na\"{\i}ve: $c=50, r=2$     & 0.06110       & 1368      & 559       & 0.04446       & 10,879     & 3170       & 0.04161       & 1,023,897  & 256,776   \\  
\hspace*{0.5cm}Na\"{\i}ve: $c=50, r=6$     & 0.03833       & 961       & 484       & 0.02199       & 5946       & 2305       & 0.01862       & 504,703    & 172,502   \\
\hspace*{0.5cm}Na\"{\i}ve: $c=200, r=6$    & 0.05872       & 1376      & 920       & 0.01180       & 3238       & 1957       & 0.00121       & 32,842     & 12,810    \\
\hspace*{0.5cm}Na\"{\i}ve: $c=500, r=6$    & 0.05994       & 1382      & 971       & 0.01319       & 3463       & 2156       & 0.00106       & 31,242     & 12,323    \\
\hline
Actions 10                            &               &           &           &               &            &            &               &            &            \\
\hspace*{0.5cm}CMABFAS                & {\bf 0.04831} & {\bf 1616}& {\bf 397} & {\bf 0.01141} & {\bf 5576} & {\bf 864}  & {\bf 0.00070} & 99,757     & {\bf 1312} \\
\hspace*{0.5cm}Na\"{\i}ve: $c=50, r=2$     & 0.13012       & 3228      & 1276      & 0.0579        & 20,330     & 5,715      & 0.04238       & 1,725,953  & 428,724    \\
\hspace*{0.5cm}Na\"{\i}ve: $c=50, r=6$     & 0.10599       & 2901      & 971       & 0.03428       & 10,225     & 3013       & 0.01747       & 500,767    & 147,629    \\
\hspace*{0.5cm}Na\"{\i}ve: $c=200, r=6$    & 0.15095       & 3378      & 1329      & 0.04826       & 16,967     & 4354       & 0.00342       & 107,174    & 23,170     \\
\hspace*{0.5cm}Na\"{\i}ve: $c=500, r=6$    & 0.15343       & 3413      & 1387      & 0.05233       & 17,769     & 4993       & 0.00272       &{\bf 98,568}& 19,958     \\
\hline
Actions 50                            &               &           &           &               &            &            &               &             &         \\
\hspace*{0.5cm}CMABFAS                & {\bf 0.08067} & {\bf 4124}& {\bf 2570}& {\bf 0.01745} &{\bf 13,921}&{\bf 9917}  & {\bf 0.00175} &{\bf 252,437}& 217,973 \\
\hspace*{0.5cm}Na\"{\i}ve: $c=50, r=2$     & 0.3153        & 8473      & 6190      & 0.12692       & 49,595     & 28,147     & 0.01367       & 796,852     & 353,873 \\
\hspace*{0.5cm}Na\"{\i}ve: $c=50, r=6$     & 0.26239       & 7785      & 5416      & 0.10556       & 42,182     & 20,700     & 0.01208       & 736,471     & 304,461 \\
\hspace*{0.5cm}Na\"{\i}ve: $c=200, r=6$    & 0.30579       & 7965      & 5541      & 0.13792       & 46,887     & 25,165     & 0.01569       & 762,208     & {\bf 190,428} \\
\hspace*{0.5cm}Na\"{\i}ve: $c=500, r=6$    & 0.30810       & 8008      & 5576      & 0.14247       & 46,899     & 25,258     & 0.01685       & 788,416     & 209,656\\
\hline
\end{tabular}
\end{center}
\caption{Quantitative results of the SPIT filter. We compare CMABFAS with a na\"{\i}ve baseline implementation for various settings of its hyperparameters (see text). The table shows the performance at three different points in time: at the beginning of the learning (after 10,000
calls), towards the middle of the learning (after 100,000 calls), and towards the end of learning (after 10,000,000 calls). Performance is given
in terms of regret/t, where regret is equal to the sum of the expected reward obtained when we would have chosen the best action minus expected
reward of the action that our SPIT filter has chosen (thus zero regret means we have always chosen the best action). The column nmistakes1 shows
the number of times the SPIT filter chooses an action which is not optimal with respect to the expected reward (an error which can mean, for example, that the SPIT filter correctly chose to apply a security challenge to SPIT call but not the security challenge with the highest probability of success/least cost ratio). The column nmistakes2 shows the number of times the SPIT filter 
chooses action A0 for a SPIT call (i.e., fails to block SPIT) or chooses one of actions A1$\ldots$A50 for a NON-SPIT call (i.e., applies
a security challenge to a NON-SPIT call). The best result for each case is marked in bold face; we can see that in terms of regret CMABFAS is about an
order of magnitude better than the best baseline.}
\label{tab:results}
\end{sidewaystable}

%==================================================================
\section{Conclusion}
%==================================================================

In this paper we have undertaken first steps towards making a complex decision-making SPIT filter
(i.e., a SPIT filter which has to choose among more than two alternatives without access to prior
labeled data and only based on stochastic and sparse feedback) become fully self-learning by formulating
it as a contextual multi-armed bandit. The simulation results are encouraging; it should be noted though
that due to the nature of the problem  
(SPIT is largely hypothetical and barely existent nowadays but believed to be a potential threat as VoIP becomes 
more widespread in the future)
our results are biased
by the modeling decisions we had to make (e.g., setting success probabilities by hand). Nevertheless, 
we believe that this research is both highly innovative and useful and could also be applied to other 
security-related problems which can be formulated in a similar way.

%ACKNOWLEDGMENTS are optional
\section*{Acknowledgments}
{\small  Sylvain Martin acknowledges the financial support of the Belgian National Fund of Scientific Research (FNRS). 
Tobias Jung acknowledges financial support from a ULg research fellowship.
This work is also partially funded by {EU} project {R}esume{N}et, FP7--224619.
}

\bibliographystyle{abbrv}
\bibliography{main}  
\end{document}